\documentclass[12pt]{iopart}
\usepackage{iopams}
\expandafter\let\csname equation*\endcsname\relax
\expandafter\let\csname endequation*\endcsname\relax
\usepackage{amsmath}
\usepackage{graphicx}
\usepackage{hyperref}
\usepackage{cite}
\usepackage{braket}
\usepackage{overpic}

\newcommand{\Haml}{\ensuremath{H_{\mathrm{latt}}}}
\newcommand{\Hamq}{\ensuremath{H_{\mathrm{qubit}}}}
\newcommand{\Hamint}{\ensuremath{H_{\mathrm{int}}}}
\newcommand{\Had}{\ensuremath{U_{\mathrm{Had}}}}
\newcommand{\up}{\ensuremath{\uparrow}}
\newcommand{\down}{\ensuremath{\downarrow}}
\newcommand{\sigmaz}{\ensuremath{\sigma_z}}
\newcommand{\hnlatt}{\ensuremath{\hat{n}_\mathrm{latt}}}
\newcommand{\nlatt}{\ensuremath{n_\mathrm{latt}}}
\newcommand{\rhoq}{\ensuremath{\rho_{\mathrm{qubit}}}}
\newcommand{\rhol}{\ensuremath{\rho_{\mathrm{latt}}}}
\newcommand{\xq}{\ensuremath{x_{\mathrm{pr}}}}
\newcommand{\omq}{\ensuremath{\omega_{\mathrm{qubit}}}}

\newcommand{\tfin}{\ensuremath{t_{\mathrm{fin}}}}

\begin{document}

\title{Quantum probe spectroscopy for cold atomic systems}

\author{%
A. Usui$^1$, 
B. Bu\v{c}a$^2$, and 
J. Mur-Petit$^2$
}
\address{%
$^1$ Quantum Systems Unit, OIST Graduate University, Onna, Okinawa 904-0495, Japan\\
$^2$ Clarendon Laboratory, University of Oxford, Parks Road,
  	Oxford OX1 3PU, United Kingdom}
\eads{\mailto{ayaka.usui@oist.jp}, \mailto{berislav.buca@physics.ox.ac.uk}, \mailto{jordi.murpetit@physics.ox.ac.uk}}

\begin{abstract}
We study a two-level impurity coupled locally to a quantum gas on an optical lattice.
For state-dependent interactions between the impurity and the gas,
we show that its evolution encodes information on the local excitation spectrum of gas at the coupling site.
Based on this, we design a nondestructive method to probe the system's excitations in a broad range of energies by measuring the state of the probe using standard atom optics methods.
We illustrate our findings with numerical simulations for quantum lattice systems, including realistic dephasing noise on the quantum probe,
and discuss practical limits on the probe dephasing rate to fully resolve both regular and chaotic spectra.
\end{abstract}



\date{\today}
\maketitle


\section{Introduction}

Atomic gases trapped in optical lattices offer unique opportunities for quantum simulation of strongly-correlated phases of matter~\cite{Bloch2012,Johnson2014} as recently demonstrated with the observation of antiferromagnetic correlations in the ground state of Hubbard-model quantum simulators~\cite{Hart2015,Cocchi2016,Parsons2016,Boll2016,Cheuk2016,Drewes2017}.
A powerful tool to study these systems are quantum gas microscopes~\cite{Hart2015,Cocchi2016,Parsons2016,Boll2016,Cheuk2016,Drewes2017,Bakr2009,Sherson2010}, that permit high-fidelity control and measurement of atoms with single-site resolution with laser fields by implementing high-resolution optical imaging systems.
A complementary experimental approach especially suitable to study transport properties is the scanning gate microscope recently developed at ETH~\cite{Hausler2017}.
Still, in analogy to the wide variety of experimental techniques available to study condensed matter systems,  it is necessary to develop a range of techniques to characterise a quantum simulator, probing for instance its density, multi-particle correlations, temperature, or excitation spectrum.

Most tools currently available for these tasks rely either on the interaction of the trapped atoms with laser fields or on density measurements after a period of expansion. 
As a classical example of a light-based technique, Bragg spectroscopy was developed in early cold atoms experiments to observe the low-energy excitation spectrum of atomic gases~\cite{Kozuma1999,Stenger1999,Rey2005,Veeravalli2008}, a method more recently employed to map the band structure of bosonic superfluids in optical lattices~\cite{Ernst2010}.
The excitation spectrum of atomic gases has also been probed by stimulated Raman spectroscopy~\cite{Dao2007,Stewart2008,Gaebler2010}, which is akin to angle-resolved photoemission spectroscopy (ARPES) in condensed matter physics~\cite{Damascelli2004}.
Nondestructive probing of atomic ensembles in cavities by analysing their interaction with quantum light has also been discussed, e.g., in Refs.~\cite{Mekhov2007,Caballero-Benitez2015njp}.
Regarding methods that exploit the wave nature of the atomic field, noise interferometry~\cite{Altman2004,Folling2005,Greiner2005} and matter-wave interferometry~\cite{Cronin2009}, which require the release of the atoms from the trap, have been successfully used to determine local and nonlocal density correlations in quantum gases.

The progress in control and measurement methods at the single-atom level enables an alternative approach based on utilising quantum impurities (e.g., single atoms in a different internal state or belonging to an entirely distinct atomic species) as nondestructive quantum probes of many-body quantum systems~\cite{Recati2005,Bruderer2006,Kollath2007STM,
Zipkes2010,Schmid2010,Will2011,Hunn2012,Spethmann2012,Fukuhara2013,
Haikka2013a,Mayer2014,Elliott2016,Streif2016,Cosco2017,%
Sabin2014,Correa2015,Johnson2016thermo,Hangleiter2015,Mitchison2016,
Hohmann2016,Hohmann2017,Schmidt2018,
Bentine2017
}.
For example, Ref.~\cite{Kollath2007STM} described a scanning tunnelling microscope analogue for atomic systems based on a single strongly-localised impurity atom, capable of measuring the local density and density correlations with nanometer resolution.
More recently, Refs.~\cite{Elliott2016,Streif2016} have proposed protocols to measure nonlocal particle correlations in atomic gases utilising one~\cite{Elliott2016} or multiple impurities~\cite{Streif2016}.
Conversely, Ref.~\cite{Cosco2017} has shown how a Bose-Hubbard lattice can act as a controllable environment leading to either Markovian or non-Markovian evolution of an impurity coupled to it.

Hangleiter \textit{et al.}~\cite{Hangleiter2015} have discussed a method to measure the excitation spectrum of a quantum gas by coupling it to an atomic impurity in a double-well potential. By tuning the parameters of the double well, they showed that the impurity's dynamical evolution becomes sensitive to the phononic excitations with energy and momentum selectivity.
Nondestructive probing of the system's dynamic structure factor using an anharmonically trapped impurity has been discussed in~\cite{Mitchison2016}.

These various theoretical proposals have accompanied by considerable experimental progress. 
Ref.~\cite{Hohmann2016} has reported temperature measurements based on monitoring the evolution of a small number of caesium impurities interacting with an ultracold rubidium gas in an optical trap, which has further enabled to study the relaxation of non-thermal states at the level of single atomic collisions~\cite{Hohmann2017}.
More recently, the coherent internal (spin) evolution of atomic impurities immersed in a condensate has been observed with high temporal ($\lesssim\:$ms) and spatial ($\sim\mu$m) resolution~\cite{Schmidt2018}, demonstrating the possibility to use the former as local quantum probes of a complex quantum gas.
In an alternative experimental approach, Refs.~\cite{Bentine2017,Harte2018} have developed the trapping of different rubidium isotopes in highly-tunable multiple radiofrequency traps.

Here, we propose a protocol to measure a broad range of the excitation energies of a quantum gas simultaneously, by coupling it to a localised two-level impurity.
Specifically, we show that monitoring the internal dynamics of the impurity enables to robustly detect small energy gaps, $\Delta E \ll J$ (with $J$ the characteristic energy scale of the system), over a broad energy range in the system's spectrum, with the lower resolution limit set by the probe dephasing rate.
Thus, our protocol constitutes a new tool to characterise cold-atom systems in optical lattices.

The paper is organised as follows. In Sec.~\ref{sec:model} we describe the model of the lattice system under consideration, and how we couple a quantum probe to it.
We provide an analytic description of the evolution of the probe in Sec.~\ref{sec:analytical-res}.
In Sec.~\ref{sec:numerical-res}, we compare the analytic results with exact numerical simulations of the protocol, considering two scenarios for the quantum probe: isolated or subject to dephasing.
Finally, we summarise our findings and discuss the relation of our proposal with earlier works in Sec.~\ref{sec:disc}. For clarity, some details of the derivation are presented in three Appendices.


\section{Description of model and protocol} \label{sec:model}
\subsection{Model setup} \label{ssec:model}
Let us consider a tight-binding model in a finite lattice with $L$ sites and $N$ particles. This system can be described by the Hamiltonian 
\begin{equation}
  \Haml
  =
  \sum_{ \langle l,m \rangle } J_{l,m} c^{\dagger}_l c_m
  + \sum_l \epsilon_l c^{\dagger}_l c_l
  \:,
  \label{eq:Hlatt}
\end{equation}
where $J_{l,m}$ represents the hopping rate between (nearest-neighbour) sites $m$ and $l$,
$\epsilon_l$ defines a single-site energy term, and 
$c_l,c_l^{\dagger}$ are the particle annihilation and creation operators at site $l$.
This model can represent a variety of experimental setups, including cold atoms in optical lattices~\cite{Bloch2012}, arrays of superconducting circuits~\cite{Underwood2012,Underwood2016}, photonic waveguides~\cite{OBrien2009},
microwave cavity arrays~\cite{Poli2015}, and optomechanical setups~\cite{Aspelmeyer2014}.

The spectral properties of this simple Hamiltonian depend sensitively on the shape of the system, and can show regular or chaotic features%
~\cite{StockmannBook}.
For instance, Ref.~\cite{Fernandez-Hurtado2014} showed that the model Eq.~\eqref{eq:Hlatt} with $\epsilon_l=0$ on a square lattice in a rectangular $L_x\times L_y$ domain present a regular spectrum, 
with a Poisson distribution of energy gaps, $P_{\mathrm{Poisson}}(s) = \exp(-s)$, with $s$ the suitably normalised energy-level spacing~\cite{Fernandez-Hurtado2014}.
On the other hand, the same model on a Bunimovich stadium  [cf.\ inset in Fig.~\ref{fig:signal-stadium}(b)] has a chaotic spectrum, which is characterised by level repulsion, i.e., no two levels are close in the energy spectrum~\cite{StockmannBook}.
This flexibility renders model~\eqref{eq:Hlatt} a useful test-bed to assess the resolution in energy of a spectroscopy protocol.

In addition, the transport dynamics on these finite lattices is relatively insensitive to the differing spectral statistics~\cite{Fernandez-Hurtado2014}, an effect that can be related to a symmetry of the square lattice~\cite{Mur-Petit2014}.
This contrasting behaviour between spectral and transport properties of finite lattices 
makes probing directly their spectrum in a manner complementary to transport measurements~\cite{Hausler2017} an interesting task in itself.

\begin{figure}[tb]
 \begin{center}
 \includegraphics[width=8cm]{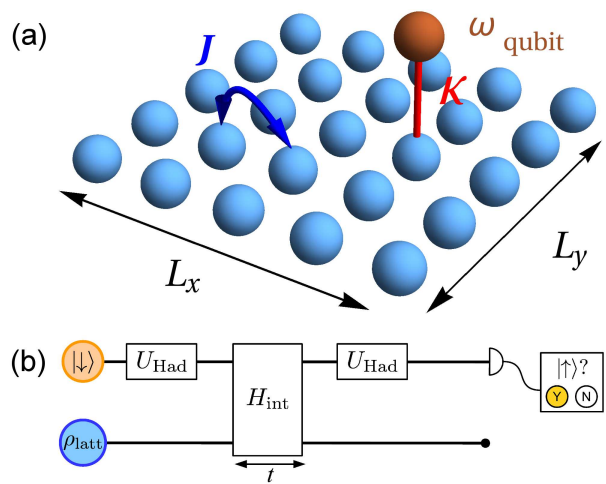}
 \end{center}
 \caption{%
 (a) Sketch of the system: Particles can hop at a rate $J$ (blue arrow) between nearest-neighbour sites on a lattice of $L_x \times L_y$ sites (light blue spheres). A quantum probe (dark orange sphere) is coupled locally to the lattice site $\xq$ with strength $\kappa$ (red line).
 (b) Probing protocol: the probe is initialised in its ground state, $\ket{\down}$, and follows a Ramsey sequence, interacting with the lattice for a time $t$ before being measured in the $\{ \ket{\up},\ket{\down} \}$ basis.
 }
 \label{fig:sketch}
\end{figure}

Our probing protocol (described below) relies on the accumulation of a differential phase between the two states of the probe by their coupling to the lattice system. 
Ref.~\cite{Haikka2013a} has shown that a probe formed by two internal states of a strongly localised atom, as the atomic quantum dot described in~\cite{Kollath2007STM,Recati2005,Fedichev2003}, is notably more susceptible to dephasing when coupled to a Bose-condensed gas than a probe comprised by an impurity in a double-well potential. 
To increase our protocol sensitivity, we thus choose to 
couple the system~\eqref{eq:Hlatt} to a localised two-level quantum system (a qubit), with internal states $\ket{\up},\ket{\downarrow}$ separated by an energy gap $\hbar\omq$. 
The corresponding Hamiltonian reads
\begin{equation}
 \Hamq=\frac{1}{2}\hbar\omq\sigmaz
 \label{eq:Hqubit}
\end{equation}
with the Pauli $z$ matrix $\sigmaz=\ket{\up}\bra{\up}-\ket{\down}\bra{\down}$.
The probe will be coupled locally to a single lattice site, located in position $\xq$, see Fig.~\ref{fig:sketch}.

%
%
The composite system is then described by the Hamiltonian
\begin{equation}
  H = 1_\mathrm{qubit}\otimes\Haml + \Hamq\otimes 1_\mathrm{latt} + \Hamint \:.
\end{equation}
We consider a state-dependent contact coupling between the lattice and the qubit of the form%
~\cite{Haikka2013a,Streif2016}
\begin{equation}
 \Hamint
 =
 \left( \kappa_{\up} \ket{\up}\bra{\up} + \kappa_{\down}\ket{\down}\bra{\down} \right)
 \otimes \hnlatt(\xq) \:.
 \label{eq:Hint}
\end{equation}
This interaction Hamiltonian describes that each internal state of the probe couples with different strength to the total density, $\hnlatt(x_i)=c^\dagger_{i} c_{i}$, at site $\xq$.
For instance, in a cold-atom implementation, this can be realised by exploiting a magnetic of laser-induced Feshbach resonance in the collisions between the atoms in the lattice and the probe. Without loss of generality, below we set $\kappa_{\up}=\kappa$, $\kappa_{\down}=0$; further, for simplicity we also set $J_{l,m} = J$.

\subsection{Quantum probing protocol} \label{ssec:protocol}

For weak coupling $\kappa \ll \{ J, \omq \}$, in accordance with perturbation theory, the interaction Hamiltonian~\eqref{eq:Hint} induces a shift of the qubit energy eigenstates,
\begin{equation}
 E_{\up,\down} \mapsto E_{\up,\down} + \kappa_{\up,\down} \nlatt \:,
\end{equation}
where $\nlatt = \mathrm{Tr}[\rhol \hnlatt(\xq)] $ is the density at the site probed.
If the qubit is prepared in a pure state, it is possible to measure $\nlatt$ by monitoring the time evolution of the population in each internal state of the qubit~\cite{Kollath2007STM}.
More generally, as we presently show, it is also possible to extract information on the lattice's spectrum. To this end, we consider the following protocol [cf.~Fig.~\ref{fig:sketch}(b)]:
\begin{enumerate}
\item Initialise the probe in its ground state, $\ket{\down}$. The composite system is initially uncorrelated, $\rho(t=0) = \rhoq \otimes \rhol$, with $\rhoq = \ket{\down}\bra{\down}$, and $\rhol$ the lattice state.
\item Apply a Hadamard gate to the qubit,
  \begin{eqnarray*}
	\Had = \frac{1}{\sqrt{2}}
	  \left( \begin{tabular}{cc} 1 & 1 \\ 1 & -1 \end{tabular} \right) \:,
  \end{eqnarray*}
  in the basis $\{ \ket{\down}, \ket{\up} \}$, so that it is now in state $(\ket{\down} + \ket{\up} ) /\sqrt{2}$. As explained in Appendix~\ref{sec:derivation}, this equal superposition is favoured to extract time-dependence of the lattice dynamics maximally.
\item At time $t=0$, couple the probe to the lattice with $\Hamint$, and let it evolve for a time $\tfin$. For concreteness, we set $\kappa_{\up}(t)=\kappa$, $\kappa_{\down}(t)=0$, for $0 < t < \tfin$. (Physically, during this stage of the protocol
the two states of the qubit acquire different phases,
$ \phi_s = t E_s/\hbar$ $(s \in  \{\down, \up\})$,
due to their interaction with the particles in the lattice.)
\item At time $t=\tfin$, apply a new Hadamard gate, and finally measure the probe state in the $\{ \down, \up \}$ basis.
\end{enumerate}


\section{Analytic time evolution} \label{sec:analytical-res}

As in a standard Ramsey sequence, the last step of the protocol transforms the relative phases accumulated by the $\up,\down$ states of the qubit into different populations of the probe states. More specifically, one can determine analytically the time evolution of the composite system through the protocol by solving the von Neumann equation for the density matrix 
\begin{equation} \label{eq:heisenbergeq}
 i \hbar \frac{\partial }{\partial t} \rho = [H, \rho] \:,
\end{equation}
with the initial density matrix corresponding to an uncoupled situation, $\rho(t=0)=\rhoq\otimes\rhol$. By tracing out the lattice, one can extract the time evolution of the qubit, $\rhoq(t)=\mathrm{Tr}_\mathrm{latt}\rho(t)$.
We find that the state of the probe at the end of the protocol is described by (see Appendix~\ref{sec:derivation} for details of the derivation)
\begin{eqnarray}
 \rhoq(\tfin)
 & = &
 \begin{pmatrix}
  \rhoq^{\up\up} & \rhoq^{\up\down} \\
 \rhoq^{\down\up} &  \rhoq^{\down\down}
 \end{pmatrix}
=
 \frac{1}{2}
 \begin{pmatrix}
  1+\cos\theta(\tfin) & i\sin\theta(\tfin) \\
 -i\sin\theta(\tfin) & 1-\cos\theta(\tfin)
 \end{pmatrix} 
 \label{eq:solution_qubit}
\end{eqnarray}
with
$\theta(t)
= \tilde{\omega} t
 + \sum_{\alpha<\sigma} [ c_{\alpha\sigma} \sin\Omega_{\alpha\sigma} t
                       + d_{\alpha\sigma} ( \cos\Omega_{\alpha\sigma} t -1) ]$.
Here the summation runs over all pairs of lattice eigenstates occupying site $\xq$, and 
$\Omega_{\alpha\sigma} = \omega_{\alpha}-\omega_{\sigma}$ is the difference between the corresponding eigenenergies.
The parameters $c_{\alpha\sigma}, d_{\alpha\sigma}$ are functions of $\Omega_{\alpha\sigma}$, the local density $\nlatt$, and the relative phase between eigenstates at $x=\xq$ (see Eqs.~\eqref{eq:tomega}-\eqref{eq:d}
in Appendix~\ref{sec:derivation}).
If there are no particles at the coupled site, $c_{\alpha\sigma}=d_{\alpha\sigma}=0$, and $\tilde{\omega} = \omq$; in this case, Eq.~\eqref{eq:solution_qubit} recovers the free evolution of the probe.
If only one lattice state is present at $\xq$, again $c_{\alpha\sigma}=d_{\alpha\sigma}=0$, but $\tilde{\omega}=\omq + \kappa\nlatt/\hbar$, in agreement with the energy shift expected in perturbation theory.
In this case, monitoring the probe's time evolution allows to determine the density at the lattice site through measurements of $\tilde{\omega}$. However, one cannot access the energy of this single lattice eigenstate.

When the probe site is occupied by several eigenstates, however, an analysis of the time evolution of the population of any of the probe states,
\begin{equation}
  P_s(\tfin) = \bra{s} \rhoq(\tfin) \ket{s} = \rhoq^{ss}, \quad s \in \{ \down, \up \}
\end{equation}
allows to retrieve the spacings between lattice energy levels, $\Omega_{\alpha\sigma}$, of states present at $\xq$. To show this, we focus on the case that there are no degenerate eigenstates; we discuss briefly the degenerate case in Appendix \ref{sec:degeneracy}.

For simplicity, let us first consider the case that only two lattice states are present at $\xq$, so that there is only one non-zero frequency difference, $\Omega_{21}=\omega_2-\omega_1$.
Then, the time dependence of the probe state follows Eq.~\eqref{eq:solution_qubit} with an angle $\theta(t)$ given by
\begin{align}
\label{eq:expansion_cos}
\cos\theta(t) 
&=
\cos\left[\tilde{\omega} t+c_1 \sin\Omega_{21} t +d_1 (\cos\Omega_{21} t-1) \right] \nonumber \\
&=
\cos \left(\tilde{\omega}t -d_1\right)
	\big\{ \cos(c_1\sin\Omega_{21} t)\cos(d_1\cos\Omega_{21} t) \nonumber \\
& 		\qquad -\sin(c_1\sin\Omega_{21} t)\sin(d_1\cos\Omega_{21} t) \big\} \nonumber \\
&-\sin \left(\tilde{\omega}t-d_1 \right)
	\big\{\sin(c_1\sin\Omega_{21} t)\cos(d_1\cos\Omega_{21} t) \nonumber \\
&		\qquad +\cos(c_1\sin\Omega_{21} t)\sin(d_1\cos\Omega_{21} t) \big\}
\end{align}
By using the Jacobi-Anger expansion (cf.~Appendix~\ref{sec:jacobi-anger}), it follows that $\cos\theta(t)$ has frequency components $\tilde{\omega} \pm m \Omega_{21}$, with 
$\Omega_{21}$ the difference in energy of the two states, and $m=0, 1, \ldots$
It is straightforward to generalise this to the case of an arbitrary number of lattice states, in which case the time evolution of the probe will have components at the frequencies $\tilde{\omega}\pm m\Omega_{\alpha\sigma}$, with $m=0,1,\ldots$, and $\alpha,\sigma$ running over all pairs of lattice states.
Physically, the situation is analogous to coupling the internal state of a trapped ion (described as a two-level system, as the probe here) to its motional states in the trap (their role played here by \textit{pairs} of lattice eigenstates): the new qubit eigenfrequencies $\tilde{\omega}\pm m\Omega_{\alpha\sigma}$ ($m>0$) are analogous to a trapped ion's motional sidebands~\cite{Haffner2008}.


\section{Numerical results} \label{sec:numerical-res}

\subsection{Non-dephasing quantum probe} \label{ssec:non-dephasing}

We have performed numerical simulations to determine the capability of our protocol to study finite lattices, both either regular or chaotic spectra, and compared the results with the analytic findings in the previous section.
We set as our energy unit the hopping amplitude $J_{l,m}=J=1$, and choose a small interaction strength $\kappa<1$, so that we can compare with the analytic results from perturbation theory.

\begin{figure}[tb]
 \centering
 \includegraphics[width=\columnwidth]{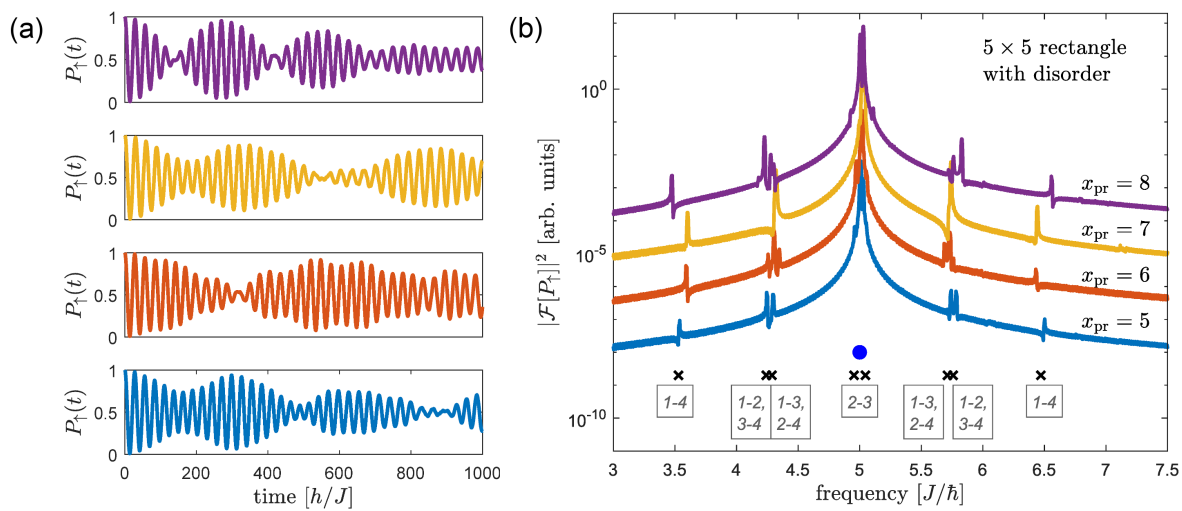}
 \caption{%
  (a) Population of the excited state of the qubit as a function of time, $P_{\up}(t)$, when it is coupled to the lattice site $\xq=8,7,6,5$ (from top to bottom) of a square lattice on a $5\times5$ rectangular domain with diagonal disorder.
  (b) Fourier transform of the signals in (a), displaced vertically for clarity with the same ordering.
 The full circle at the bottom indicates $\omq$, while the crosses are the expected frequencies $\tilde{\omega}\pm\Omega_{\alpha\sigma}$, with the states $\alpha,\sigma\in	\{\textit{1,\ldots,4}\}$ indicated in the boxes. 
 In these simulations, $\omq=5.0$ and $\kappa=0.3$,
 with the hopping rate as our energy unit, $J=1$.}
 \label{fig:signal-rectangle}
\end{figure}

We first consider a $5\times5$ square lattice on a rectangular domain, with small diagonal disorder, modelled by single site energies taken from a uniform random distribution $\epsilon_k \in [-0.3,0.3]$ (this allows to lift level degeneracies due to the high symmetry of the square lattice), and set the qubit level splitting to $\omq=5$.
To test our protocol, we take as the lattice initial a superposition of the four lower-energy eigenstates, that we label \textit{1,2,3,4} (our protocol is likewise applicable when the lattice system is in a mixed state, cf.~Appendix~\ref{sec:derivation}).
We then expect the time evolution of the probe to show six first-order ($m=1$) sidebands in frequency space, with varying amplitudes depending on the coupling site.
We show in Fig.~\ref{fig:signal-rectangle}(a) the time evolution of the excited state population of the probe when coupled to various lattices sites, $\xq$. These time traces show a complex behaviour, which is easier to understand by moving to frequency space.

The Fourier transform of these signals is reported in Fig.~\ref{fig:signal-rectangle}(b), where we can clearly identify a small number of underlying frequency components.
There is a dominant peak at $\tilde{\omega}\simeq\omq$: as $\kappa\nlatt$ is rather small ($\nlatt\lesssim0.1$), the frequency shift $\tilde{\omega}-\omq$ is unobservable at the energy resolution of the figure.
There are 12 additional narrow peaks, distributed symmetrically to lower and higher frequencies. The frequencies of all peaks are consistent with the expected locations of the first-order sidebands, $\tilde{\omega}\pm m\Omega_{\alpha\sigma}$, with $\alpha,\sigma \in \{\textit{1,\ldots,4}\}$, and it is easy to identify all peaks with pairs of lattice states%
\footnote{Taking into account that, within the frequency resolution in Fig.~\ref{fig:signal-rectangle}, $\delta\omega\simeq 10^{-1}J/\hbar$, the peaks around $\omega/(J/\hbar)=4.2, 4.3, 5.6$, and 5.7 correspond to two transitions each.}.
We do not observe peaks from higher-order sidebands, $m\geq 2$. This is due to the amplitude of each peak being proportional to a Bessel function $J_m(c_{\alpha\sigma})$ [or $J_m(d_{\alpha\sigma})$], with $c_{\alpha\sigma},d_{\alpha\sigma} \propto \kappa\nlatt$. In the present simulations, we have $\kappa\nlatt \lesssim 10^{-2}$. In this limit, $J_m(x) \lesssim 10^{-4}$ for $m\geq 2$, which is below the resolution in Fig.~\ref{fig:signal-rectangle}.
(We discuss in Appendix~\ref{sec:stft} practical requirements on measurement time to
achieve the required frequency resolution in light of typical parameters in current experiments.)

An important observation of Fig.~\ref{fig:signal-rectangle}(b) is the variation in the number of frequency peaks, as well as in their locations and intensities, as the coupling site is modified. For instance, when the probe is coupled to site $\xq=5$, there are two distinct peaks at $\omega \approx 4.3$. When the probe is displaced to $\xq=6$, there are three similarly intense peaks, while for $\xq=7$, we see one large peak only. These variations spring from the spatial dependence of the various eigenstates. This is also reflected for instance in the displacement in frequency of the peak at $\omega\approx3.5$ depending on the probe position.
These findings support that our protocol is able to capture the different energy spacings in the spectrum of a generic lattice system in a position-dependent way, from which the local density of states can be reconstructed.

\begin{figure}[!tb]
 \centering
 \includegraphics[width=\columnwidth]{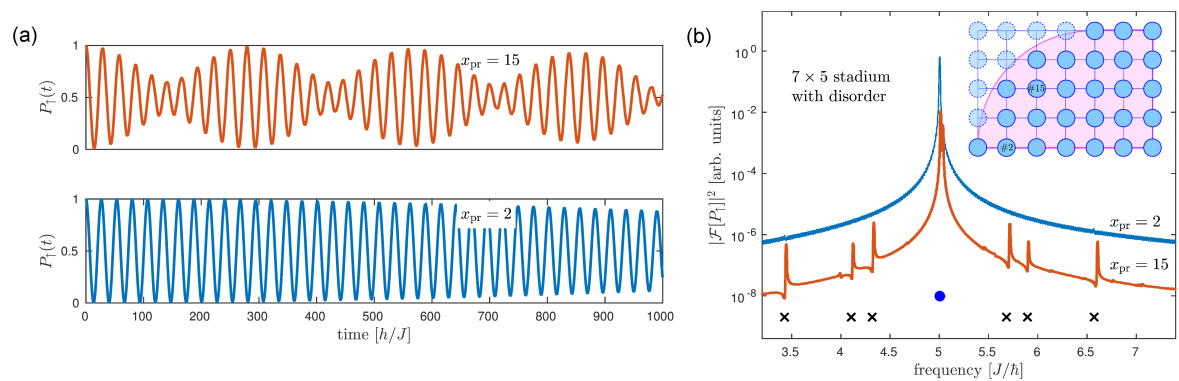}
 \caption{%
 (a) Population of the excited state of the qubit as a function of time, when coupled to an occupied (top panel, $\xq=15$) or empty (bottom, $\xq=2$) site of a square lattice on a $7\times5$ Bunimovich stadium.
 (b) Fourier transform of the signals in (a).
 The crosses at the bottom indicate the expected frequencies $\tilde{\omega}\pm\Omega_{\alpha\sigma}$ while the full circle is at $\omq$.
 (Inset) Scheme of the finite lattice with labels $\#2,\#15$ at the sites where the probe is coupled.
 Other parameters as in Fig.~\ref{fig:signal-rectangle}.
 }
 \label{fig:signal-stadium}
\end{figure}

To further illustrate this point, we have simulated as well the time evolution of a qubit probe coupled to a square lattice on a domain with the shape of a Bunimovich stadium with 27 sites [cf.~Fig.~\ref{fig:signal-stadium}(b,inset)], for which the spectrum is chaotic~\cite{Fernandez-Hurtado2014}.
In this case, we have taken as the lattice initial a superposition of three lattice eigenstates with different energies, so that again we expect to observe six first-order sideband peak on each side of $\tilde{\omega}$.


We show in Fig.~\ref{fig:signal-stadium}(a) the time evolution of $P_{\up}$ for the case that the probe is coupled to a site populated by all seven states ($\xq=15$). The Fourier transform of this signal is reported as a thick solid line in Fig.~\ref{fig:signal-stadium}(b). As was the case with the rectangular domain, we can clearly identify each frequency component with the expected peak at $\tilde{\omega}\pm\Omega_{\alpha\sigma}$ ($\alpha,\sigma\in\{ \textit{1,2,3} \}$), which illustrates the power of the protocol to unravel rather complicated energy spectra.
Additionally, in this case we observe a small displacement of $\tilde{\omega}-\omq=0.06$, for the signal taken at $\xq=15$, which agrees with the perturbation theory expectation with peak density $\nlatt\approx0.2$.
As a final check, we also coupled the qubit to a site that is not populated by any of the lattice states ($\xq=2$, bottom panel in Fig.~\ref{fig:signal-stadium}(a)]). The corresponding Fourier signal [top line in Fig.~\ref{fig:signal-stadium}(b)] features only the peak at $\omq$ as predicted by Eq.~\eqref{eq:tomega} in this case.


\subsection{Effect of dephasing on the quantum probe} \label{ssec:dephasing}


\begin{figure}[tb]
 \centering
  \includegraphics[width=\columnwidth]{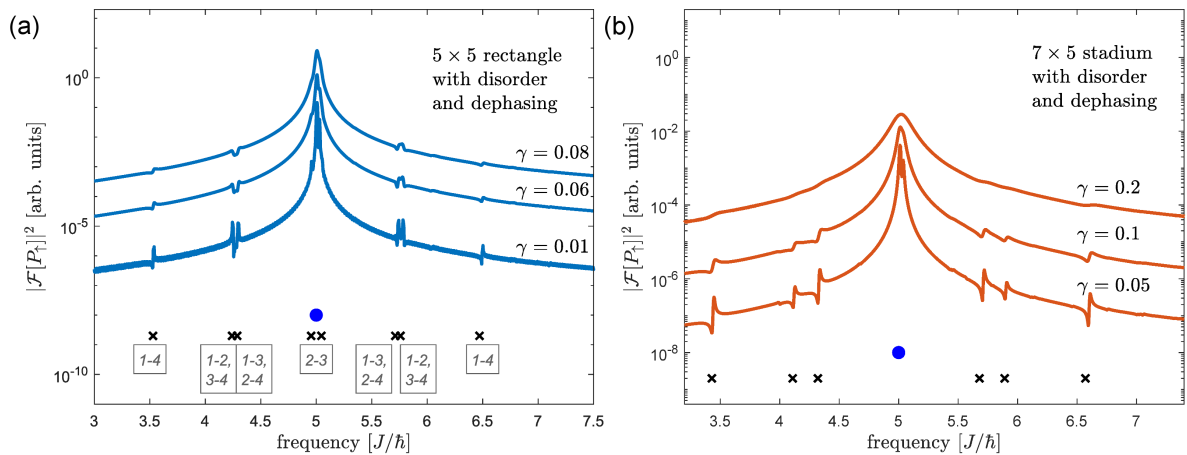}
 \caption{(color online)
 (a) Fourier transform of the signal for a probe coupled to site $\#5$ of the $5\times5$ rectangle with disorder, and subject to dephasing with various dephasing rates $\gamma = 0.01, 0.06, 0.08$ as indicated;  other parameters and symbols as in Fig.~\ref{fig:signal-rectangle}. The different traces are displaced vertically for clarity.
 (b) Fourier transform of the signal for a probe coupled to site $\#15$ of the $7\times5$ stadium with disorder, and subject to dephasing with various dephasing rates $\gamma = 0.05, 0.1, 0.2$ as indicated;  other parameters and symbols as in Fig.~\ref{fig:signal-stadium}.
 In both panels, the different traces are displaced vertically for clarity.}
	\label{fig:dephasing}
\end{figure}

A real probe will be inevitably coupled to the environment, and it is important to assess to what extent the accuracy of the idealised measurements simulated in Sect.~\ref{sec:numerical-res}.A is affected by this.
In an implementation in which the lattice is realised with cold atoms in an optical lattice, and the quantum probe by an atom trapped in a separate optical lattice or in optical tweezers, we expect dephasing of the internal state of the probe due to the trapping and ambient electromagnetic fields to be the dominant source of noise~\cite{Kuhr2005}.
This can be modelled using the standard Markovian approach to open quantum systems~\cite{BreuerBook}.
In this formalism, the evolution of the density matrix describing the lattice and probe is described by the Lindblad master equation
\begin{equation}
\frac{\partial}{\partial t}\rho(t)
= 
-i [H,\rho(t)] + \gamma \left(2L \rho(t) L^\dagger - \{L^\dagger L,\rho(t)\}\right) 
\label{eq:lindeq}
\end{equation}
where the Lindblad operator $L=\sigmaz$ dephases the probe in the $z$-direction, and $\gamma$ is the dephasing rate.

We present in Fig.~\ref{fig:dephasing} simulations of the joint evolution of the lattice and probe including dephasing noise according to Eq.~\eqref{eq:lindeq} for the rectangle and the stadium.
(The calculation was done via a Suzuki-Trotter decomposition of the Liouvillian; this allows us to work with operators rather than superoperators, greatly increasing the efficiency of the numerical calculation, see~\cite{Rivas2012ch4}.)

Fig.~\ref{fig:dephasing}(a) shows the Fourier transform of the signal for a probe coupled to a $5\times5$ rectangle with disorder, and subject to dephasing with various dephasing rates $\gamma = 0.01, 0.06, 0.08$. For very small dephasing rate, $\gamma \ll 1$, there is no noticeable effect. On the other hand, as expected, when $\gamma$ becomes comparable to the distance between the peaks, they merge and can no longer be distinguished; see for instance the merging of the two peaks around $\omega\approx5.7$ for $\gamma \gtrsim 0.06$. For $\gamma > 0.08$ practically all peaks have become unobservable.

Interestingly, the level repulsion between eigenstates in chaotic systems renders the measurement of their energy gaps with the present protocol more robust against probe dephasing. This is illustrated in Fig.~\ref{fig:dephasing}(b), where we present the Fourier transform of the probe signal for a disordered $7\times5$ Bunimovich stadium, including dephasing noise on the probe.
Here, as the dephasing rate $\gamma$ is increased, the peaks loose strength in a similar way to what is found for the rectangle. However, the absence of nearby pairs of peaks makes it possible to resolve the various dominant frequencies up to larger dephasing rates, $\gamma \approx 0.1$: each peak eventually becomes unobservable for strong dephasing, but they do not get to merge, in contrast that what is found in the rectangle. 
In general, the Fourier peaks for both regular and chaotic spectra will be distinguishable if one can control the dephasing rate of the probe below a threshold which may depend on the kind of lattice. (We emphasise that, within the approach embodied by Eq.~\eqref{eq:lindeq}, there is no energy exchange between the probe and the source of environmental noise, which guarantees that the peaks in the Fourier signal are not displaced; this would not hold in the presence of dissipative noise.)


\section{Discussion} \label{sec:disc}

In summary, we have studied the dynamics of a two-level quantum probe locally coupled to a quantum lattice system.
We have shown that the probe's evolution encodes information on the local density and excitation spectrum of the lattice system, and designed a nondestructive protocol to measure them based on state-dependent probe-system contact interactions and standard control and measurement techniques applied on the probe. 
Our numerical simulations including dephasing of the probe support the applicability of our protocol to study lattices with either regular or chaotic spectra.

The key ingredients of our proposal are a two-level probe on which we only require projective measurements in the computational basis ($\{\ket{\uparrow},\ket{\downarrow}\}$), and a local density-density coupling to the system of interest.
The simple level structure of the probe makes monitoring its dynamics easier than for the case of probes realised with a quantum harmonic oscillator, as recently proposed in Ref.~\cite{Nokkala2016} to measure the spectral density of a large structured environment (i.e., in the limit $N\to\infty$), which requires to measure the average excitation number of the probe.

The density-density coupling to the system makes our protocol sensitive to the presence of particles at the coupling site, and thus readily applicable to bosonic or fermionic many-body lattice systems.
On the other hand, a probe with a richer structure ---together with a more complex coupling to the system--- would be required to perform full counting statistics of particle occupations on the sites that would reveal the quantum statistics of the system.

Our probing strategy is nondestructive, essentially encoding the system's excitations into the probe's phase, which is then accessed by a Ramsey sequence with measurements in the $\{\ket{\uparrow},\ket{\downarrow}\}$ basis. This strategy sets lower experimental requirements than more elaborate protocols aimed at determining the structure or internal couplings of spin networks~\cite{Burgarth2009njp,Burgarth2011,Kato2014}, which ask for full state tomography.


Finally, our method does not rely on a resonant coupling between the probe and the system~\cite{Hangleiter2015}, thus enabling one to measure various spectral gaps simultaneously, even if the impurity is subject to additional dephasing processes. Because of these reduced requirements, our protocol constitutes an attractive tool to characterise the spectrum of systems implemented with cold atoms in optical lattices.
We expect this work will contribute to the development of new measurement techniques~\cite{Kollath2007STM,Hunn2012,Mayer2014,Elliott2016,Streif2016,Cosco2017,%
Sabin2014,Correa2015,Johnson2016thermo,Hangleiter2015,Mitchison2016}
exploiting atomic impurities to characterise cold-atom quantum simulators~\cite{Spethmann2012,Hohmann2016,Hohmann2017,Schmidt2018,Bentine2017,Harte2018}, and to explore aspects of quantum chaos in ultracold finite-sized systems~\cite{Fernandez-Hurtado2014,Mur-Petit2014}.


\ack
 We acknowledge useful discussions with E. Bentine and Th.\ Busch.
 This work was supported by Okinawa Institute of Science and Technology Graduate University,
 EPSRC Grants Nos.\ EP/P009565/1 and EP/P01058X/1,
 the EU H2020 Collaborative project QuProCS (Grant Agreement 641277),
 and Spain's MINECO Grant No.~FIS2015-70856-P. 


\appendix
\section{Time evolution of the qubit} \label{sec:derivation}
We start with the von Neumann Eq.~(\ref{eq:heisenbergeq}), which we write explicitly in terms of the matrix elements,
\begin{equation}
 \label{eq:heq_elements}
 i \hbar \frac{\partial }{\partial t} \rho_{s,\alpha;k,\beta}
 =
 \sum_{r,\sigma}\left[
 	H_{s,\alpha;r,\sigma}\rho_{r,\sigma;k,\beta}
 	-\rho_{s,\alpha;r,\sigma}H_{r,\sigma;k,\beta} \right] \:
\end{equation}
Here, $\rho_{s,\alpha;k,\beta}= \langle s,\alpha|\rho|k,\beta\rangle$ are the elements of the density matrix, while $H_{s,\alpha;k,\beta}= \langle s,\alpha|H|k,\beta\rangle$; we use Roman indices $s,k,r$ to refer to probe eigenstates, and Greek indices $\alpha,\beta,\sigma$ for lattice states. By tracing out the lattice states, the left hand side (l.h.s) of Eq.~\eqref{eq:heq_elements} can be recast in the form
\begin{align}
 \sum_{\alpha,\beta}
 & (\mathrm{l.h.s.})
   \delta_{\alpha,\beta}
=
 \left(
 	i \hbar \frac{\partial \tilde{\rho}_{s,k}}{\partial t}
 	+ \hbar \left(\omega_s-\omega_k\right) \tilde{\rho}_{s,k} \right)
   e^{-i(\omega_s-\omega_k)t}
 \label{eq:LH}
\end{align}
with the probe eigenenergies $\omega_{s(k)}$ and $\rho_{s,k}(t)=\tilde{\rho}_{s,k}(t) e^{-i(\omega_s-\omega_k)t}$.

In general, the qubit and lattice become entangled by the interaction. However, we make an assumption that density matrix is separable at all times, $\rho(t) = \rhoq(t) \otimes \rhol(t)$, which permits to simplify some matrix elements in Eq.~\eqref{eq:heq_elements}:
$\bra{s,\alpha} (1_{\mathrm{qubit}}\otimes\Haml) \rho \ket{k,\beta}
=\braket{s|\rhoq|k}\bra{\alpha} \Haml \rhol \ket{\beta}$
and
$\bra{s,\alpha} (\Hamq\otimes1_{\mathrm{latt}}) \rho \ket{k,\beta}
=\braket{s| \Hamq \rhoq |k} \braket{\alpha|\rhol |\beta}$.
This assumption is rigorously justified for weak coupling and short evolution times, but our numerical results support a broader applicability for the present problem.

Tracing over the lattice states on the right hand side (r.h.s) of Eq.~\eqref{eq:heq_elements}, we then obtain
\begin{align}
 \sum_{\alpha,\beta}
  & (\mathrm{r.h.s.})
    \delta_{\alpha,\beta}
 =
 \hbar\left(\omega_s-\omega_k\right)\rho_{s,k} 
  + \sum_{\alpha,r,\sigma} \left(
 	\Hamint^{s,\alpha;r,\sigma} \rho_{r,\sigma;k,\alpha}
 	-\rho_{s,\alpha;r,\sigma} \Hamint^{r,\sigma;k,\alpha}  \right) \:.
 \label{eq:RH}
\end{align}
For the contact interaction Eq.~\eqref{eq:Hint}, the matrix elements of the interaction Hamiltonian 
are
$\Hamint^{s,\alpha;k,\beta}
= \bra{s,\alpha} \Hamint \ket{k,\beta} 
= \kappa \delta_{s,k}\delta_{s,\up} \braket{\alpha|\xq} \braket{\xq|\beta}$,
with $\braket{\xq|\beta}$ the amplitude of lattice eigenstate $\ket{\beta}$ at site $\xq$, and $\braket{\alpha|\xq}=\braket{\xq|\alpha}^*$.
We substitute this result in Eq.~\eqref{eq:RH}, apply the separability assumption again, and finally combine with Eq.~\eqref{eq:LH} to rewrite Eq.~\eqref{eq:heq_elements} as
\begin{equation}
 \label{eq:heq_qubit}
 i \hbar \frac{\partial}{\partial t}
 \begin{pmatrix}
  \tilde{\rho}_{\up\up} & \tilde{\rho}_{\up\down} \\
  \tilde{\rho}_{\down\up} & \tilde{\rho}_{\down\down}
 \end{pmatrix}
 =
 \begin{pmatrix}
  0 & M(t)\tilde{\rho}_{\up\down} \\
  -M(t)\tilde{\rho}_{\down\up} & 0
 \end{pmatrix},
\end{equation}
with
\begin{equation}
 \label{eq:M}
 M(t) = \sum_{\alpha} \kappa A_{\alpha\alpha}
  	   +\sum_{\alpha<\sigma} 2 \kappa A_{\alpha\sigma}
  	   \cos \left\{ (\omega_\alpha-\omega_\sigma)t+\phi_{\alpha\sigma} \right\} .
\end{equation}
Here, we introduced
$A_{\alpha\sigma}e^{i\phi_{\alpha\sigma}}=\langle \alpha|\xq \rangle \langle \xq|\sigma \rangle\tilde{\rho}_{\sigma\alpha}$ with real numbers $A_{\alpha\sigma}>0$ and $\phi_{\alpha\sigma}$.
The first summation in Eq.~\eqref{eq:M} runs over all lattice eigenstates,
while the second runs over all pairs of eigenstates.
Physically, the factors $\braket{\alpha|\xq},\braket{\xq|\sigma}$ guarantee that only eigenstates with nonzero probability density at $\xq$ contribute to the evolution of the probe's off-diagonal terms.
On the other hand, importantly, in this derivation the lattice initial state does not need to be a pure state, but it can be a general mixed density matrix, which implies that our method can be applied likewise to quantum gases with a nonzero thermal component~\cite{Goold2011}.

From Eq.~\eqref{eq:heq_qubit} it follows that only the off-diagonal elements evolve, in accordance with the fact that the interaction Hamiltonian describes a dephasing of the probe state. This requires the initial state to have non-zero off-diagonal elements; the optimal choice is an equal-weight superposition such as $(\ket{\down} + \ket{\up} ) /\sqrt{2}$, cf.~\cite{Benedetti2018}.
At the end of the evolution and after the final Hadamard gate, the state of the qubit is of the form Eq.~\eqref{eq:solution_qubit} with
\begin{align}
 \label{eq:theta}
 \theta(t)
 =& 	\left(\omq + \sum_{\alpha} \frac{\kappa A_{\alpha\alpha}}{\hbar}  \right) t
+
  \sum_{\alpha<\sigma} \frac{2\kappa A_{\alpha\sigma}}{\hbar\Omega_{\alpha,\sigma}}
   \left[\sin \left( \Omega_{\alpha,\sigma}t+\phi_{\alpha\sigma} \right)
   -\sin \phi_{\alpha\sigma} \right] .
\end{align}
This has the form given in the main text,
$\theta(t)
= \tilde{\omega} t
 + \sum_{\alpha<\sigma} [ c_{\alpha,\sigma} \sin\Omega_{\alpha,\sigma} t
                       + d_{\alpha,\sigma} ( \cos\Omega_{\alpha,\sigma} t -1) ]$,
with
\begin{align}
 \tilde{\omega}
  &=  \omq + \frac{\kappa}{\hbar} \sum_{\alpha} A_{\alpha \alpha}
 \:, \label{eq:tomega} \\  
 \Omega_{\alpha,\sigma} &= \omega_\alpha - \omega_\sigma
 \:, \qquad \qquad
 \eta_{\alpha,\sigma} = \frac{2\kappa A_{\alpha \sigma}}{\hbar\Omega_{\alpha,\sigma}}
 \:, \label{eq:Omegan} \\
 c_{\alpha,\sigma} &= \eta_{\alpha,\sigma} \cos(\phi_{\alpha \sigma})
 \:, \qquad 
 d_{\alpha,\sigma} = \eta_{\alpha,\sigma} \sin(\phi_{\alpha \sigma})
 \:.
 \label{eq:d}
\end{align}
We have solved numerically the von Neumann equation~\eqref{eq:heq_elements} with the initial state $\rho_{s,k}(t=0)=1/2$ $\forall s,k\in \{\up,\down\}$. As shown in Figs.~\ref{fig:signal-rectangle}-\ref{fig:signal-stadium}, 
the numerical results of Eq.~\eqref{eq:heq_elements} agree with the analytic results~\eqref{eq:theta}, which justifies the separability assumption.


\section{Case of energy degeneracy} \label{sec:degeneracy}
Consider a lattice system with energy degeneracy between eigenstate $s_1$ and $s_2$. The dynamics of the probe follows Eq. (\ref{eq:heq_qubit}), with $M(t)$ given by 
\begin{align}
M(t)
=
 \sum_{\alpha} \kappa A_{\alpha\alpha} + 2 \kappa A_{s_1 s_2} \cos \phi_{s_1 s_2} 
+
 \sum_{\substack{\alpha<\sigma\\ \alpha,\sigma\neq s_1,s_2}} 2 \kappa A_{\alpha\sigma}
 \cos \left\{ (\omega_\alpha-\omega_\sigma)t+\phi_{\alpha\sigma} \right\},
\end{align}
which shows that an extra term is added into Eq.(\ref{eq:M}). This means that level degeneracy leads to changes in the frequency $\tilde{\omega}$, but does not disturb observation of lattice energy levels $\Omega_{\alpha,\sigma}$.

\section{Jacobi-Anger expansion} \label{sec:jacobi-anger}
For completeness, we include here explicit expressions of the Jacobi-Anger expansion relevant to Eq.~\eqref{eq:expansion_cos}, cf.~\cite{AbramowitzBook}:
\begin{eqnarray}
 \!\!\!  \!\!\! 
 \cos(z \cos \phi) &=& J_0(z) + 2\sum_{k=1}^\infty (-1)^k J_{2k}(z)\cos(2k\phi) \\
  \!\!\!  \!\!\! 
 \cos(z \sin \phi) &=& J_0(z) + 2\sum_{k=1}^\infty J_{2k}(z)\cos(2k\phi) \\
  \!\!\!  \!\!\! 
 \sin(z \cos \phi) &=& 2\sum_{k=0}^\infty (-1)^k J_{2k+1}(z)\cos[(2k+1)\phi] \\
  \!\!\!  \!\!\! 
 \sin(z \sin \phi) &=& 2\sum_{k=0}^\infty J_{2k+1}(z)\sin[(2k+1)\phi]
 \label{eq:jacobi-anger}
\end{eqnarray}
with $J_k(z)$ the Bessel function of 1st kind and order $k$.

\section{Measurement time and frequency uncertainty} \label{sec:stft}

It is of practical importance to assess how long one needs to monitor the qubit probe in order to retrieve spectral information on the system, particularly when the probe is subject to large dephasing rates ($\gamma \gtrsim 0.1 J$).
The trade-off between frequency and observation time that follows from the Fourier transform is encapsulated in the Wiener-Heisenberg relation between angular frequency resolution $\Delta\omega$ and measurement time $\tfin$~\cite{Gabor1946,Boashash2003}
\begin{equation*}
  \tfin\Delta\omega \geq 1/2.
\end{equation*}
For cold atoms in optical lattices, one has typical hopping rates $J/\hbar \sim 1-100$ Hz.
Typical system lifetimes are limited by vacuum to $\sim 10-70$~s~\cite{Hohmann2015,Endres2016}, which would enable to resolve peaks down to $\Delta\omega\gtrsim 10^{-2}-10^{-1}$~Hz.
This appears sufficient to discern the peaks in the most demanding situation in our simulations: nearby peaks in the disordered rectangle are separated by $\simeq10^{-2} J/\hbar$, which corresponds to $\sim 10^{-2}-1$~Hz, depending on $J$. 
Still, the longer lifetime of impurities immersed in a quantum gas reported to date is 40~ms~\cite{Mayer2018}, with prospects of increasing up to $\simeq 600$~ms~\cite{Hohmann2015}.

\section*{References}
\bibliographystyle{iopart-num}

\providecommand{\newblock}{}

\end{document}